\documentclass[a4paper]{spie}  

\usepackage{amsmath,amsfonts,amssymb}
\usepackage{graphicx}
\usepackage[colorlinks=true, allcolors=blue]{hyperref}

\title{RISTRETTO: reflected-light exoplanet spectroscopy at the diffraction limit of the VLT}

\author[a,b]{Christophe Lovis}
\author[a]{Nicolas Blind}
\author[a]{Bruno Chazelas}
\author[a]{Muskan Shinde}
\author[a]{Maddalena Bugatti}
\author[a]{Nathana\"el Restori}
\author[a]{Isaac Dinis}
\author[a]{Ludovic Genolet}
\author[a]{Ian Hughes}
\author[a]{Micha\"el Sordet}
\author[a]{Robin Schnell}
\author[a]{Samuel Rihs}
\author[a]{Adrien Crausaz}
\author[c,d]{Martin Turbet}
\author[a]{Nicolas Billot}
\author[e,f]{Thierry Fusco}
\author[f]{Benoit Neichel}
\author[e,f]{Jean-Fran\c cois Sauvage}
\author[a]{Pablo Santos Diaz}
\author[a]{Mathilde Houelle}
\author[g]{Joshua Blackman}
\author[a]{Audrey Lanotte}
\author[g]{Jonas K\"uhn}
\author[a]{Janis Hagelberg}
\author[h,i]{Olivier Guyon}
\author[j]{Patrice Martinez}
\author[j]{Alain Spang}
\author[g]{Christoph Mordasini}
\author[a,b]{David Ehrenreich}
\author[g]{Brice-Olivier Demory}
\author[a,b]{Emeline Bolmont}

\affil[a]{D\'epartement d'Astronomie, Universit\'e de Gen\`eve, CH-1290 Versoix, Switzerland}
\affil[b]{Centre pour la Vie dans l'Univers, Universit\'e de Gen\`eve, Geneva, Switzerland}
\affil[c]{Laboratoire de M\'et\'eorologie Dynamique, IPSL, CNRS, Sorbonne Universit\'e, 4 place Jussieu, F-75252 Paris Cedex 05, France}
\affil[d]{Laboratoire d'Astrophysique de Bordeaux, Univ. Bordeaux, CNRS, B18N, all\'ee Geoffroy Saint-Hilaire, 33615 Pessac, France}
\affil[e]{DOTA, ONERA, Universit\'e Paris Saclay, F-91123 Palaiseau, France}
\affil[f]{Laboratoire d'Astrophysique de Marseille, CNRS, CNES, Aix Marseille Universit\'e, 38 rue Fr\'ed\'eric Joliot Curie, F-13013 Marseille, France}
\affil[g]{Weltraumforschung und Planetologie, Physikalisches Institut, Universit\"at Bern, Gesellschaftsstrasse 6, CH-3012 Bern, Switzerland}
\affil[h]{Subaru Telescope, National Astronomical Observatory of Japan, National Institutes of Natural Sciences (NINS), 650 North Aohoku Place, Hilo, HI 96720, USA}
\affil[i]{Steward Observatory, University of Arizona, Tucson, AZ 87521, USA}
\affil[j]{Universit\'e C\^ote d'Azur, CNRS, Laboratoire Lagrange, Nice, France}

\authorinfo{Send correspondence to C.L. E-mail: christophe.lovis@unige.ch}

\begin{document} 
\maketitle

\begin{abstract}
RISTRETTO is a visible high-resolution spectrograph fed by an extreme adaptive optics (AO) system, to be proposed as a visitor instrument on ESO VLT. The main science goal of RISTRETTO is to pioneer the detection and atmospheric characterisation of exoplanets in reflected light, in particular the temperate rocky planet Proxima b. RISTRETTO will be able to measure albedos and detect atmospheric features in a number of exoplanets orbiting nearby stars for the first time. It will do so by combining a high-contrast AO system working at the diffraction limit of the telescope to a high-resolution spectrograph, via a 7-spaxel integral-field unit (IFU) feeding single-mode fibers. Further science cases for RISTRETTO include the study of accreting protoplanets such as PDS70b/c through spectrally-resolved H-alpha emission, and spatially-resolved studies of Solar System objects such as icy moons and the ice giants Uranus and Neptune. The project is in the manufacturing phase for the spectrograph sub-system, and the preliminary design phase for the AO front-end. Specific developments for RISTRETTO include a novel coronagraphic IFU combining a phase-induced amplitude apodizer (PIAA) to a 3D-printed microlens array feeding a bundle of single-mode fibers. It also features an XAO system with a dual wavefront sensor aiming at high robustness and sensitivity, including to pupil fragmentation. RISTRETTO is a pathfinder instrument in view of similar developments at the ELT, in particular the SCAO-IFU mode of ELT-ANDES and the future ELT-PCS instrument.
\end{abstract}


\section{INTRODUCTION}
\label{sec:intro}  

In 2016, the discovery of Proxima b clearly marked a milestone in the exoplanet field: the star closest to the Sun, at just 1.3 pc, hosts a temperate rocky planet in its habitable zone\cite{Anglada2016,Suarez2020,Faria2022}. Given its proximity, angular separation and favorable contrast with respect to the star, there will never be a "better" potentially habitable planet than this one in terms of observability. This triggered the original idea to develop RISTRETTO, a pioneering experiment for reflected-light exoplanet spectroscopy at the VLT\cite{Lovis2017,Lovis2022}. Combining an XAO system to a high-resolution spectrograph in the visible, we could show that Proxima b would be amenable to direct detection with an 8m telescope. RISTRETTO would thus become a pathfinder for this science case, which requires an ensemble of state-of-the-art but existing technologies to be combined together for the first time. This would pave the way towards the development of similar instrumentation at the ELT, namely ANDES and PCS. Technical specifications for RISTRETTO were derived from the requirement to characterize Proxima b, which serves as the sizing science case for the instrument. However they broadly apply to reflected-light exoplanet spectroscopy in general. They can be summarized as follows: an inner working angle of at most 40 mas, a planet coupling efficiency of $>$50\%, a stellar coupling (contrast) of $<$10$^{-4}$, a spectral resolution $R>$ 100,000, a spectral range $\Delta\lambda/\lambda$ $>$ 25\%, and a total system throughput $>$5\%. We thus started to design an instrument that would be made of three essential parts: a high-Strehl XAO system, a coronagraphic integral-field unit working at the diffraction limit, and a visible high-resolution spectrograph. We also developed a  simulator that allows us to explore the feasibility of the various RISTRETTO science cases, which are briefly described in this paper.

\section{INSTRUMENT OVERVIEW}

The main sub-systems of the RISTRETTO instrument are listed below. We refer to the corresponding papers presented at this conference for details on the status of each sub-system.

\begin{itemize}

\item Ultra-fast, high-Strehl XAO system: the need to efficiently couple the planet light into the spectrograph calls for achieving a Strehl ratio $>$70\% in $I$-band. Moreover, the need to achieve a raw contrast of 10$^{-4}$ close to the diffraction limit calls for a very fast correction loop ($>$2 kHz) to minimize servo-lag error. The current XAO baseline design foresees a dual wavefront sensor made of an $H$-band unmodulated Pyramid for high sensitivity to low-order modes, complemented by a $zY$-band Zernike wavefront sensor for optimized sensing of phase discontinuities across the pupil (e.g. low-wind effect). Both WFS will share a common detector, a fast C-RED ONE near-IR camera. Wavefront correction will be achieved by a woofer-tweeter system comprising an ALPAO 97-15 woofer with 100 actuators and a Boston Micromachines 2K-3.5 tweeter with 2000 actuators. We refer to papers 13097-213 (Blind et al.), 13097-189 (Shinde et al.), 13097-295 (Shinde et al.), and 13097-192 (Motte et al.) for detailed end-to-end simulations of the RISTRETTO XAO system and test results from the C-RED ONE camera.

\item Coronagraphic IFU: RISTRETTO will use the newly-developed concept of the PIAA nuller\cite{Blind2022} (PIAA-N) to feed a bundle of 7 single-mode fibers arranged in an hexagonal pattern covering the 2-$\lambda/D$ annulus around the central star. This solution offers both a high throughput for planets located within this annulus and strong suppression of the stellar PSF.  We refer to paper 13100-103 (Restori et al.) for a detailed description of this setup and the presentation of test results from a PIAA-N prototype.

\item Visible high-resolution spectrograph: the RISTRETTO spectrograph will cover the 620-840 nm wavelength range at a spectral resolution $R$=140,000. It is a diffraction-limited design accepting 7 single-mode fibers as input\cite{Chazelas2022}. The spectrograph is currently in the manufacturing and procurement phase, with assembly, integration and tests planned for 2024-2025. Tests of the vacuum vessel and thermal enclosure are ongoing to demonstrate the required thermo-mechanical stability of the instrument once installed on the VLT Nasmyth platform. We refer to paper 13096-283 (Chazelas et al.) for a detailed status update of this sub-system.

\item Instrument simulator: we are developing a full end-to-end simulator for RISTRETTO as a tool to prepare the scientific observing programme. It includes realistic XAO and PIAA performances and uses the Pyechelle software package\cite{Sturmer2019} to generate simulated raw echelle spectra. Performances on real star-planet systems such as Proxima Centauri are being investigated and advanced data analysis techniques are being developed to optimally extract the planetary signal. We refer to paper 13096-358 (Bugatti et al.) for a detailed description of this tool.

\end{itemize}

\section{EXOPLANET TARGET LIST FOR RISTRETTO}

\begin{figure} [ht]
\begin{center}
\includegraphics[width=\textwidth]{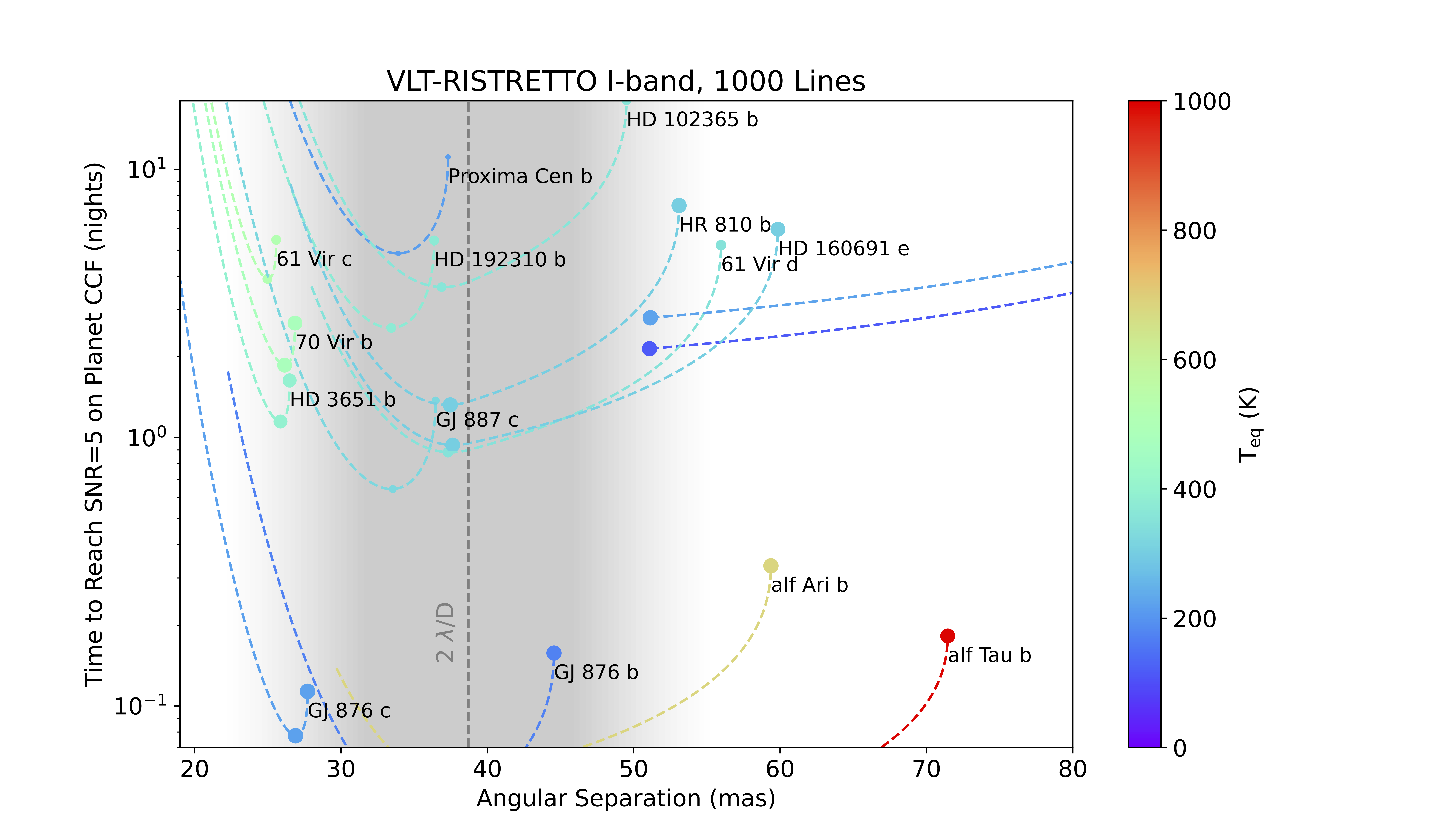}
\end{center}
\caption[example] 
{ \label{fig:target_list} 
Required exposure time to reach SNR=5 on the planet cross-correlation signal as a function of angular separation from the star. The simulation assumes 1000 spectral lines (reflected stellar lines or planetary lines) and an albedo of 0.4. Dashed lines for each planet show the exposure times at different orbital phases assuming an orbital inclination of 60 degrees and a Lambertian phase curve. The shaded area around 2 $\lambda/D$ shows the planet coupling efficiency into the off-axis spaxels of RISTRETTO.}
\end{figure}

Over the past three decades, RV surveys have discovered many exoplanets orbiting very nearby stars. Some of them reach angular separations at maximum elongation that are large enough to make them resolvable by 8m-class telescopes in the visible. We used the NASA Exoplanet Archive and the RISTRETTO simulator to derive a target list of exoplanets that RISTRETTO will be able to characterize. As can be seen from Fig.~\ref{fig:target_list}, a dozen of known objects are accessible to RISTRETTO, from cold Jupiters to warm Neptunes to the temperate rocky planet Proxima b. The cold gas giants GJ876b and GJ876c can be detected in just 1-2 hours of observing time. The warm Saturn HD3651b can be detected in about 1 night of integration. The Neptune-mass objects HD192310b, HD102365b and 61 Vir d can be studied in a few nights. The warm super-Earth GJ887c can be reached in less than a night. And Proxima b, if observed at its brightest accessible phase angle, can be detected in about 5 nights if it has an Earth-like atmosphere. For all these objects, the level of characterization will depend on a priori unknown planet properties such as albedo and atmospheric composition. Broadband and chromatic albedos may be derived from the RISTRETTO detection of the reflected stellar spectral lines. Phase curves may be obtained for planets that can be observed at different phase angles. And molecular absorption by water vapor, methane and oxygen may be detected if present in sufficient quantities. Thus RISTRETTO is by no means a "single-target instrument" but truly has the capability to open a new field in exoplanet studies.

\section{OTHER SCIENCE CASES}

Besides exoplanets in reflected light, RISTRETTO will address a variety of other science cases that will be of broad interest to the planetary science and stellar communities. We briefly summarize them here:

\begin{itemize}

\item Accreting protoplanets in H-alpha: the RISTRETTO wavelength range covers the H-alpha line, which is a well-known tracer of accretion in young stars and protoplanets. Thus we will be able to study the spectrally-resolved H-alpha line profile in objects such as PDS70b/c and constrain the geometry and kinematics of the accretion flows onto the planet. Moreover, RISTRETTO will be able to search for new protoplanets at 30-50 mas from young stars, which corresponds to $\sim$3-5 AU at the distance of nearby star-forming regions. This is the typical orbital distance at which most giant planets are thought to form, and thus RISTRETTO will be able to access this crucial parameter space for the first time.

\item Kinematics of protoplanetary disks: RISTRETTO will be able to derive spatially-resolved Doppler velocity maps of protoplanetary disks in scattered light. This capability would nicely complement the sub-mm velocity maps obtained by ALMA. Resolved disk kinematics including localized velocity anomalies linked to forming planets could be probed with RISTRETTO.

\item Spatially-resolved stellar surfaces: with an angular resolution of 19 mas, RISTRETTO will be able to resolve the surfaces of the largest stars in the sky, e.g. Betelgeuse (45 mas in diameter) or Antares (41 mas). It will thus be able to study how spectral lines vary in shape and radial velocity as a function of disk position, and how they evolve with time.

\item Solar System science: RISTRETTO will be able to study local structures at the surfaces of the icy moons of Jupiter and Saturn (e.g. Io, Europa, Titan) with a spatial resolution of 120 km at Jupiter and 240 km at Saturn. It may also measure local wind velocities and chemical abundances in the atmospheres of Uranus and Neptune.

\end{itemize}

\section{RELEVANCE AND IMPACT}

This project will demonstrate novel concepts in XAO and high-contrast spectroscopy that will find direct applications in the ELT era. At the European level, RISTRETTO joins the SPHERE upgrade project SAXO+\cite{Boccaletti2022} as one of two XAO demonstrators for the ELT. RISTRETTO will differentiate itself from SAXO+ in several respects:

\begin{itemize}
\item XAO in the visible, where exoplanets are expected to be most reflective (higher albedos), and thus a unique wavelength range that ELT-PCS will target as well
\item Inner working angle down to 2 $\lambda/D$ or 38 mas, required on both the VLT and ELT for targeting exoplanets around nearby stars in reflected light
\item Unmodulated $H$-band PyWFS for maximal sensitivity to low-order aberrations
\item Integrated dual WFS strategy for enhanced sensitivity and robustness to phase discontinuities across the pupil
\item Coronagraphic integral-field unit
\item Efficient coupling to a high-resolution spectrograph
\end{itemize}

In terms of science return, RISTRETTO is well positioned to be the first to explore Proxima b, our nearest neighbour and the most accessible temperate rocky world there will ever be. Beyond Proxima b, the RISTRETTO experiment will pioneer exoplanet reflected-light spectroscopy, which is a completely new observational approach. We will for the first time be able to directly probe the atmospheres and surfaces of the exoplanets that are closest to the Solar System. These comprise a diversity of objects, from cold Jupiters to warm Neptunes to terrestrial worlds. Their study will be highly complementary to the ongoing characterization of transiting exoplanets with e.g. JWST. RISTRETTO will offer a first glimpse into this population of nearby planets, which will be the one also targeted by ELT-ANDES, ELT-PCS, and NASA Habitable Worlds Observatory (HWO) in the future.

\acknowledgments 

This work has been carried out within the framework of the National Centre of Competence in Research PlanetS supported by the Swiss National Science Foundation under grants 51NF40\_182901 and 51NF40\_205606. The RISTRETTO project was partially funded through the SNSF FLARE programme for large infrastructures under grants 20FL21\_173604 and 20FL20\_186177. The authors acknowledge the financial support of the SNSF.

\bibliography{report} 
\bibliographystyle{spiebib} 

\end{document}